\documentstyle[12pt,epsf]{article}

\def\be{\begin{equation}}
\def\ee{\end{equation}}
\def\bea{\begin{eqnarray}}
\def\eea{\end{eqnarray}}

\begin{document}

\begin{flushright}
hep-ph/0203115
\end{flushright}

\pagestyle{plain}

\def\e{{\rm e}}
\def\haf{{\frac{1}{2}}}
\def\tr{{\rm Tr\;}}
\def\goes{\rightarrow}
\def\ie{{\it i.e.}, }
\def\tcl{T_{\rm cl}}
\def\Goes {\Rightarrow}
\def\CX{{\cal X}}
\def\cm{{\rm c.m.}}
\def\fp{{\rm f.p.}}
\def\ca{{\cal A}}
\def\cf{{\cal F}}
\def\cd{{\cal D}}
\def\cv{{\cal V}}
\def\cvsym{{\cal V}_{{\rm sym.}}}
\def\cvnonsym{{\cal V}_{{\rm non-sym.}}}
\def\iphi{{\bf i}_\Phi}
\def\iv{{\bf i_v}}
\def\xgphi{{x\goes\Phi}}

\def\bsx{{\bf x}}
\def\bbx{{\bf X}}
\def\bsp{{\bf p}}
\def\bbp{{\bf P}}
\def\bba{{\bf A}}
\def\bsv{{\bf v}}
\def\bsk{{\bf k}}

\begin{center}
\vspace{1cm}
{\Large {\bf On Relevance Of Matrix Coordinates}}\\
\vspace{.4cm}

{\Large {\bf For The Inside Of Baryons}}

\vspace{.8cm}

Amir H. Fatollahi

\vspace{.4cm}

{\it Institute for Advanced Studies in Basic Sciences (IASBS),}\\
{\it P.O.Box 45195-159, Zanjan, Iran}

\vspace{.4cm}

{\sl fath@iasbs.ac.ir}
\end{center}
\vskip .3 cm

\begin{abstract}
It is argued that one natural choice for coordinates of constituents of a baryonic
state in a SU($N$) gauge theory are $N\times N$ hermitian matrices. It is discussed
that the relevance of matrix coordinates is supported at least by the restricted form
of the color symmetry. Based on the previous investigations in this direction, the
consequences of the idea are reviewed. The model has been considered is
originated by the D0-branes of String Theory.
\end{abstract}

\vspace{7.0cm}



%12.38.Aw General properties of QCD
%14.65.-q Quarks
%02.40.Gh Noncommutative geometry

\newpage

%%%%%%%%%%%%%%%%%%%%%%%%%%%%%%%%%%%%%%%%%%%%
One of the main themes in quantum mechanics is to found our physical theories
exclusively upon relationships between quantities which in principle are observable
\cite{werner}. At the present status, it is commonly believed that a hadron has quarks
as part of its ingredients, though they cannot be detected directly. From the pure
theoretical point of view, one quark on its own is like the other particles, and has
some observable quantities, such as position, momentum, spin or charge. In practice,
seemingly we are always faced with hadrons that the properties of quarks are hidden
inside them. Although, it does not seem natural to assume that quarks do not carry any
of the usual degrees of freedom or their degrees of freedom can be completely ignored,
it may be a desirable  framework if it is possible that the degrees of freedom can
become ``unreachable" due to some kind of symmetry. In other words, due to a symmetry
it would not be expected that, for example, the position of an individual quark can be
measured, or even the question about ``the position of an individual quark with a
specific color" become meaningless.

In \cite{02414,05484,fat021,fat241}, a model was considered which shares the feature
we mentioned in above. The model has been considered is originated by the D0-branes
\cite{9510017,tasi} of String Theory, for which it is known that their degrees of freedom
are captured by matrices, rather than numbers \cite{9510135}. The concerned model in
\cite{02414}-\cite{fat241} has shown its ability to reproduce or cover some features and
expectations in hadron physics. Some of these features and expectations are:
phenomenological inter-quark potentials, the behavior of total scattering amplitudes,
rich polology of scattering amplitude, behavior in large-$N$ limit, and the whiteness of
baryons with respect to the SU($N$) sector of the external fields.

As mentioned, the internal dynamics of the D0-brane bound state is
described by a matrix model of coordinates, and if the matrix
coordinates of D0-branes have something to do with hadron physics,
it is very logical to ask ``Is it possible to extract or derive
these matrices from some first principles of quantum
chromodynamics (QCD)?" In fact the answer to this question has
been the motivation for the present work, and as we will see, it
appears that the appearance of matrix coordinates in the theory of
quarks is as natural as the appearance in the theory of D0-branes.
Before presentation the much more formal derivation, let us
present the heuristic argument. Reminding the procedure of
reasoning in D0-brane theory, we recall that the matrix
coordinates are the result of some states, to be specific some
open string states, which are equipped with two more labels than
the usual ones, the so-called Chan-Paton labels
\cite{9510135,tasi,pol-book}. In the open string picture these
labels are attached to the ends of string. On the other side,
D0-branes are defined as point-like objects to whom the open
strings end. By this picture, each D0-brane is accompanied with
some more degrees of freedom than the usual ones of an ordinary
particle. In other words, each D0-brane has some more degrees of
freedom which express to which other D0-branes and in what places
is connected, \ie has made a bound state with which others.
Eventually it appears that in a bound state of $N$ D0-branes, the
relevant degrees of freedom in each direction of space, rather
than $N$, are $N^2$ which may be represented by a matrix belonging
to the U($N$) algebra. Now as we shall recognize in a moment, this
reasoning is applicable for the case of quarks, in which the
states have the additional degrees of freedom  as ``color".

In the constituent quark picture of a SU($N$) gauge theory, a
baryon is made by $N$ quarks in different colors, and besides, to
bring the baryon state as a singlet in color space, an
anti-symmetrization in the color labels is understood. Let us for
the moment forget about the rotation in the color space, and
assume that the baryon is just made by $N$ quarks in different
colors, represented by the states and wave functions
$|\psi_a(t)\rangle$ and $\langle \bsx_a|\psi_a(t)\rangle
=\psi_a(\bsx_a,t)$, $a=1,\cdots, N$, respectively;
in this work we also do not care about the fermionic or bosonic nature
of quarks. Though the wave functions depend on different
arguments $\bsx_1,\cdots, \bsx_N$,
while keeping the right form of each function $\psi_a$, we may
present them by one argument $\bsx$, as $\psi_a(\bsx,t)$. By this
we can define the $N\times N$ matrix $\bbx$ via its elements
$\bbx_{ab}(t)\equiv \int d\bsx\; \psi^*_b(\bsx,t) \bsx
\psi_a(\bsx,t)$. Here we assume that the states are normalized
properly, to yield the length dimension for the elements of
$\bbx$, accompanied with the value one for the total probability.
It is easily seen that $\bbx$ is $N\times N$ hermitian matrix, and
its elements are characterized by the color labels $a,b=1,\cdots,
N$.

Let us take the case for which we have well separated quarks, may
be represented by wave functions $\psi_a(\bsx,t)\simeq
\delta(\bsx-\bsx_a)$ with $|\bsx_a -\bsx_b|\gg \ell$, ($a\neq b$),
for some characteristic length $\ell$. For this case, the matrix
$\bbx(t)$ is almost, or even in the case, exactly diagonal.
Suppose we take the length scale $\ell$ to be the order of the
baryon size. From our experience, we know that the situation we
have considered above is never seen in practice! The most expected
situation is that, due to confinement, the $N$ quarks find
considerable overlap between their wave-functions and form a baryon.
Correspondingly, we have learnt to face always with permanently
`connected' quarks, for which the matrix $\bbx(t)$ appears always
in non-diagonal form, and this may cause one believes in the
essence of more degrees of freedom as representatives and
also to describe the permanent connectivity of quarks.
We note that in fact the huge amount of
information about the inside of a baryon is encoded in the
wave functions of its constituents, or equivalently in the matrix
coordinate $\bbx$ and its generalizations to higher moments as
$\bbx_{ab}^{(n)}(t)\equiv \int d\bsx\; \psi^*_b(\bsx,t) \bsx^n
\psi_a(\bsx,t)$, $n=0,2,3,4,\cdots$. The above simple observation
may suggest that the matrix coordinate $\bbx$ and its
generalizations to higher moments can present the criteria for the
identification the confined phase of a theory. Besides, the
matrix coordinate $\bbx$ and its higher moments may be taken as a
set of very powerful tools for characterization and study of the
observable states in a confined theory. Therefore, it will be very
tempting to see that by considering the matrix $\bbx(t)$ as the
dynamical variable relevant for the inside of a baryon, what kind
of information and conceptual insights come out.

Before to proceed further, it is useful to mention that the
matrix coordinate can also be constructed from the original
quark field in the Lagrangian. We take a SU($N$) gauge theory,
consisting one kind of flavor in the fundamental representation as
matter. We treat this example as a quantum mechanics, rather than a field theory.
The states of matter in this quantum mechanical problem are represented by
\bea
|\Psi(t)\rangle=\left(\!\!\!
\begin{array}{c} |\psi_1(t)\rangle \\ |\psi_2(t)\rangle \\
\vdots\\ |\psi_N(t)\rangle
\end{array}\!\!\!\!\right).
\eea
So we have the expansion $|\Psi(t)\rangle=\int d\bsx
\sum_{a=1}^N \psi_a(\bsx, t) |\bsx\rangle\otimes |a\rangle$, in which the
index $a$ is labelling the isospin, and $\psi_a(\bsx,t)\equiv \langle\bsx|
\psi_a(t)\rangle$. We define the density matrix operator $\hat{\rho}(t)$ as
\bea
\hat{\rho}(t)\equiv |\Psi(t)\rangle\langle\Psi(t)|,
\eea
which is an $N\times N$ matrix with the general element as $\hat{\rho}_{ab}(t)
=|\psi_a(t)\rangle\langle\psi_b(t)|$. By the density operator $\hat{\rho}_{ab}(t)$,
we can evaluate a particular expectation value for the position operator simply by:
\bea
\bbx(t)\equiv {\rm tr}_\bsx (\hat{\bsx}\hat{\rho}(t))= \left(\!\!\!
\begin{array}{cccc}
\langle\psi_1(t)|\hat{\bsx}|\psi_1(t)\rangle &
\langle\psi_2(t)|\hat{\bsx}|\psi_1(t)\rangle &
\dots & \langle\psi_N(t)|\hat{\bsx}|\psi_1(t)\rangle\\
\langle\psi_1(t)|\hat{\bsx}|\psi_2(t)\rangle & \dots & \dots& \dots\\
\vdots & \ddots & \ddots & \vdots\\
\langle\psi_1(t)|\hat{\bsx}|\psi_N(t)\rangle & \dots & \dots &
\langle\psi_N(t)|\hat{\bsx}|\psi_N(t)\rangle
\end{array}\!\!\!\right),
\eea
in which $\hat{\bsx}$ is the usual position operator, and ${\rm tr}_\bsx$
means the integration on the volume of space, yielding
$\langle\psi_a(t)|\hat{\bsx}|\psi_b(t)\rangle = \int d\bsx\; \psi^*_a(\bsx,t) \bsx
\psi_b(\bsx,t)$. The general element is defined by $\bbx_{ab}(t)=\langle\psi_b(t)
|\hat{\bsx}|\psi_a(t)\rangle=\bbx^*_{ba}(t)$, and so the matrix coordinate $\bbx(t)$
is $N \times N$ hermitian matrix with the usual expansion in color (isospin) space
as $\bbx(t)=\sum_{a,b=1}^N \bbx_{ab}(t) |a\rangle\langle b|$. Again as
is recognized easily, the elements of the matrix coordinate
$\bbx$ are characterized by the color labels $a,b=1,\cdots, N$.

As usual, it is natural to assume that the expectation values satisfy some
classical equations of motion. Also, we expect that via the quantization of the
out-come classical theory, we end up with the original quantum theory.
In general case, one expects the classical equations can be derived
from the quantized theory, in particular by the equations of motion
for the states or wave functions. Since in the problem at hand the
quantum theory, specially in the non-perturbative regime, is too hard to solve
one may try to formulate the classical theory on some general grounds.
In our specific case naturally we are faced with a matrix model. So a
general classical action for the coordinates $\bbx(t)$ may be taken as
\bea\label{action1} S[\bbx]=\int dt \;\tr
\bigg(\frac{1}{2} m \dot{\bbx}\cdot \dot{\bbx} -
\cv(\bbx,\dot{\bbx},\bbx_{ab},\dot{\bbx}_{ab})\bigg),
\eea
where $\tr$acts on the matrix structure, and ``$\cv(\cdots)$" is for the possible potential term,
depending on matrix coordinate or velocity, or probably some of their individual elements
$\bbx_{ab}$ and $\dot{\bbx}_{ab}$. For the well separated quarks, as is mentioned
the coordinate matrix $\bbx(t)$ is almost, or even in the
case, exactly diagonal and the action (\ref{action1}) becomes
\bea\label{action2}
S[\bbx]\simeq S[\bsx_1,\cdots,\bsx_N]= \int dt \sum_{a=1}^N \bigg(\frac{1}{2} m
\dot{\bsx}_a\cdot\dot{\bsx}_a - \cdots\bigg),
\eea
in which $\bsx_a=\bbx_{aa}$. The kinetic term of the action (\ref{action2}) may be interpreted
as the kinetic term of $N$ quarks. This shows that our new tool, matrix coordinates,
consists the information we usually realize, in particular the positions and
velocities of individual quarks.

The issue of gauge symmetry of original quantum mechanical problem should be considered.
The theory we start with is invariant under the transformations:
\bea |\Psi(t)\rangle
&\goes& |\Psi'(t)\rangle = \hat{V}(\hat{\bsx},t)
|\Psi(t)\rangle,\nonumber\\
\langle\Psi(t)| &\goes& \langle\Psi'(t)| = \langle \Psi(t)|
\hat{V}^\dagger(\hat{\bsx},t) ,\nonumber\\
\hat{{\cal O}}(\hat{\bsx},\hat{\bsp},t,\partial_t) &\goes& \hat{{\cal O}}'
(\hat{\bsx},\hat{\bsp},t,\partial_t)=\hat{V}(\hat{\bsx},t) \hat{{\cal O}}
(\hat{\bsx},\hat{\bsp},t,\partial_t)\hat{V}^\dagger(\hat{\bsx},t),
\eea
for the Hamiltonian of the form ${\cal H}=\langle \Psi(t)|\hat{{\cal O}}
(\hat{\bsx},\hat{\bsp},t,\partial_t)|\Psi(t)\rangle$, and
$\hat{V}(\hat{\bsx},t)$ is an $N\times N$ unitary operator,
\ie $\hat{V}\hat{V}^\dagger=\hat{V}^\dagger\hat{V}={\bf 1}_N$.
Due to integration on space $\int d\bsx$, it might not be expected
that in a simple way all of the large symmetry above can be
recovered in the theory for matrix coordinates. Instead we
assume that the position
dependence of the $\hat{V}$ matrix is in the form of $\hat{V}(\hat{\bsx},t)
=\tilde{U}(\hat{\bsx})U(t)$, where $U(t)$ is an $N\times N$ unitary matrix,
and $\tilde{U}(\hat{\bsx})$ is a phase depending on the position operator
$\hat{\bsx}$, \ie $\tilde{U}\tilde{U}^*=1$. By this kind of transformations
we are treating the position dependence of matrix $\hat{V}$ as a U(1) group, rather
than a non-Abelian one. Later we try to present some kind of justification
for the restriction on the transformations. It can be seen that
the matrix coordinate transforms as
$\bbx(t)\goes \bbx'(t)=U(t)\bbx(t) U^\dagger(t)$. So our matrix theory, at least,
should be invariant under this kind of transformations,
\footnote{The invariance under the global transformations by
$\hat{V}(\hat{\bsx},t)=V_0$, with $V_0$ as a constant $N\times N$ unitary matrix,
requires that the action should not consist the individual elements of $\bbx$, as
we assumed in (\ref{action1}), in the first step.} and as usual
this can be done by introducing a covariant derivative. So the action
(\ref{action1}) can be rewritten as
\bea\label{action4}
S[a_t,\bbx]=\int dt \;\tr \bigg(\frac{1}{2} m D_t\bbx\cdot D_t\bbx -
\cv(\bbx,D_t\bbx)\bigg),
\eea
in which $D_t\bbx=\dot{\bbx}+i[a_t,\bbx]$, with $a_t(t)$ as the
one dimensional $N\times N$ hermitian gauge field. We see that the action is now
invariant under the transformations:
\bea\label{Xat}
\bbx&\goes& \bbx'=U \bbx U^\dagger,\nonumber\\
a_t&\goes&  a'_t=U a_t U^\dagger -i U \partial_t U^\dagger,
\eea
with $U\equiv U(t)$ as an arbitrary $N\times N$ unitary matrix;
in fact under these transformations one obtains
\bea\label{DTF}
D_t\bbx\goes D'_t \bbx'=U (D_t\bbx) U^\dagger,\;\;\;\;\;
D_tD_t\bbx\goes D'_t D'_t
\bbx'=U (D_tD_t\bbx) U^\dagger.
\eea

One may go a little more in details of the potential term. First, we assume
that the potential is linear in velocity $D_t\bbx$, appearing in the potential as
$D_t\bbx\cdot \bba(\bbx,t)$. Second, since here we have
matrices as coordinates, we can decompose the velocity independent term to
completely symmetric and non-symmetric parts in components of
$\bbx=(X^1, X^2, \cdots, X^d)$. We note that each component $X^i$ is a matrix.
The non-symmetric part can be expanded as
\bea
\cv_\frac{{\rm veloc.\;indepen.}}{{\rm non-symm.}}(\bbx)= \underbrace{X^i+[X^i,X^j]+X^i[X^j,X^k]}
-\frac{m}{4l^4}[X^i,X^j][X_i,X_j]+O(X^6),
\eea
in which the terms ``$\underbrace{\cdots}$" consist free space indices or traceless
parts. So the first surviving term is ``$-m[X^i,X^j]^2/4l^4$", with $l$ as
a proper length scale. Finally we require that the vector potential
$\bba(\bbx,t)$ is also symmetric in the components $X^i$'s. Putting these all
for the potential term, we end up with the action
\bea\label{action5}
S[a_t,\bbx]=\int dt \;\tr \bigg(\frac{1}{2} m D_t\bbx\cdot D_t\bbx
+qD_t\bbx\cdot \bba(\bbx,t) - qA_0(\bbx,t) +\frac{m}{4l^4}[X^i,X^j]^2\bigg),
\eea
in which $A_0(\bbx,t)$ is the symmetric part of velocity independent
term of potential, and $q$ plays the role of the
charge. We note that the fields $(A_0(\bbx,t),\bba(\bbx,t))$
appear as $N\times N$ hermitian matrices due to their functional dependence
on the matrix coordinate $\bbx$.
It is interesting to study the gauge symmetry of
this action. One can check easily that action (\ref{action5})
is invariant under the symmetry transformations \cite{0103262,0104210,0108198}:
\bea\label{NAT}
\bbx&\goes& \bbx'=U \bbx U^\dagger,\nonumber\\
a_t(t)&\goes& a'_t(t)=U a_t(t) U^\dagger -i
U \frac{d}{dt}U^\dagger,\nonumber\\
A_i(\bbx,t)&\goes& A'_i(\bbx',t)=
U A_i(\bbx,t)U^\dagger+iU\delta_i U^\dagger,\nonumber\\
A_0(\bbx,t)&\goes&  A'_0(\bbx',t)= U
A_0(\bbx,t)U^\dagger-iU\partial_t U^\dagger,
\eea
where $U\equiv U(\bbx,t)=\exp(i\Lambda)$ is arbitrary up to the condition that
$\Lambda(\bbx,t)$ is hermitian and totally symmetrized in the
$X^i$'s. In above, $\delta_i$ is the functional derivative
$\frac{\delta} {\delta X^i}$. \footnote{We note that though
$U(\bbx,t)$ depends on $\bbx(t)$, due to the total derivative
$\frac{d}{dt}$, $a'_t(t)$ still only depends on time. In this
sense the transformations in \cite{0103262,0104210,0108198} are
interpreted incorrectly.} We recall that, in approving the
invariance of the action, the symmetrization prescription on the
matrix coordinates plays an essential role \cite{0103262,0104210}.
It is by this symmetry transformation that we expect no
distinguished role should be identified to the (diagonal or
off-diagonal) elements of matrix coordinate. In other words, since
each of the matrix elements are not gauge invariant quantities,
they are not expected to appear as an observable final state.

The above transformations on the gauge potentials are similar to
those of non-Abelian gauge theories, and we mention that it is just the
consequence of enhancement of degrees of freedom from numbers ($\bsx$)
to matrices ($\bbx$). In other words, we are faced with a situation in
which ``the rotation of fields" is generated by ``the rotation of
coordinates" \cite{0104210}. In addition,
the case we see here may be considered as finite-$N$ version
of the relation between gauge symmetry transformations and transformations
of matrix coordinates \cite{0007023}. Despite the non-Abelian behavior of the
gauge transformations, we should note that the symmetry is still
not equivalent to a non-Abelian one. To see this, we should recall that the symmetry
transformations of,
for example a U($N$) gauge theory, is generated by $N^2$ functions of space-time,
say $\Lambda_\alpha(\bsx,t)$ ($\alpha=0,\cdots, N^2-1$), in the group element
$\exp(i\Lambda_\alpha T^\alpha))$, where $T^\alpha$'s are U($N$) generators.
Now although $U(\bbx,t)=\exp(i\Lambda(\bbx,t))$ in
(\ref{NAT}) is a unitary matrix due to its dependence on
matrix coordinate, it is constructed by
just one function $\Lambda(\bsx,t)$, after replacing coordinates by
matrices {\it i.e.}  $\bsx\goes\bbx$, under the condition of
symmetrization.
After all, it is quite natural to interpret the fields $(A_0,\bba)$
as the external gauge fields that the constituents, whose degrees of
freedom
are included in the matrix coordinate, interact with them.

The action (\ref{action5}) is known
to be the action of $N$ D0-branes of String Theory, in the background of
(RR) gauge field $(A_0(\bsx,t),\bba(\bsx,t))$, for $\bsx$ as
the ordinary coordinates \cite{9910053}.
As mentioned before, from the String Theory point of view, D0-branes
are point particles to which ends of strings are attached \cite{9510017}.
In a bound state of $N$ D0-branes, they are connected to
each other by strings stretched between them, and it can be shown that,
by counting the degrees of freedom for the oriented strings, the correct
dynamical variables describing the positions of D0-branes
are $N\times N$ hermitian matrices \cite{9510135}. By comparison, we find out that
$m$ is the mass of D0-branes and $l$ is the order of the string length.
In \cite{02414,05484,fat021,fat241} the possibility for the identification
of dynamics of D0-branes and quarks are investigated. Here we recall some of
the aspects mentioned in these papers. First of all, we see that by the
gauge transformation (\ref{NAT}), the elements of the position matrix
mix with each other, and so the interpretation of the positions for D0-branes
remains obscure. Nevertheless, we note that the concept of center-of-mass (c.m.),
here presented by the trace of the matrix coordinate is meaningful. So
the ambiguity of the positions only remains for the degrees of freedom
counting the relative positions of D0-branes and the strings stretched
between them. The equations of motion for $X^i$'s and $a_t$ by action
(\ref{action5}), ignoring the commutator potential
$[X_i,X_j]^2$, is
found to be \cite{0103262,0104210,0108198}
\bea\label{LORENZ}
&~&mD_tD_t X_i=q\bigg(E_i(\bbx,t)
+\underbrace{D_tX^jB_{ji}(\bbx,t)}\bigg),\\
\label{A0EOM}
&~&m[X_i,D_tX^i]=q[A_i(\bbx,t),X^i],
\eea
with the following definitions
\bea
\label{ELEC}
E_i(\bbx,t)&\equiv&-\delta_i A_0(\bbx,t)-\partial_t A_i(\bbx,t),\\
\label{MAGN}
B_{ji}(\bbx,t)&\equiv&-\delta_jA_i(\bbx,t)+\delta_iA_j(\bbx,t).
\eea
In (\ref{LORENZ}), the symbol $\underbrace{D_tX^j B_{ji}(\bbx,t)}$ denotes
the average over all of positions of $D_tX^j$ between the $X$'s of
$B_{ji}(\bbx,t)$. The above equations for the $X$'s are like the Lorentz
equations of motion, with the exceptions that two sides are $N\times N$
matrices, and the time derivative $\partial_t$ is replaced by its
covariant counterpart $D_t$.

The behavior of eqs. (\ref{LORENZ}) and (\ref{A0EOM}) under gauge
transformation (\ref{NAT}) can be checked. Since the action is invariant
under (\ref{NAT}), it is expected that the equations of motion change
covariantly. The left-hand side of (\ref{LORENZ}) changes to $U^\dagger
D_tD_tX U$ by (\ref{DTF}), and therefore we should find the same change
for the right-hand side. One can check that in fact this is the case
\cite{0103262,0104210,0108198}, and consequently one finds that
Eq. (\ref{MAGN}) leads to
\bea\label{gtfs}
E_i(\bbx,t) &\goes& E'_i(\bbx',t)=
U E_i(\bbx,t)U^\dagger,\nonumber\\
B_{ji}(\bbx,t) &\goes& B'_{ji}(\bbx',t)=
U B_{ji}(\bbx,t)U^\dagger
\eea
This result is consistent with the fact that $E_i$ and $B_{ji}$ are functionals
of $X$'s. We thus see that, in spite of the absence of the usual
commutator term $i[A_\mu,A_\nu]$ of non-Abelian gauge theories, in our
case the field strengths transform like non-Abelian ones. We recall that
these are all consequences of the matrix coordinates of D0-branes.
Finally by the similar reason for vanishing the second term of
(\ref{action5}), both sides of (\ref{A0EOM}) transform identically.

An equation of motion similar to (\ref{LORENZ}) is considered in
\cite{fat241,fat021} as a part of similarities between the dynamics of
D0-branes and bound states of quarks--QCD strings in a baryonic state
\cite{fat241,fat021,02414}. The point is that, the dynamics of the bound
state c.m. is not affected directly by the non-Abelian
sector of the background, \ie the c.m. is ``white" with respect to SU($N$)
sector of matrices. The c.m. coordinates and momenta are defined by:
\bea\label{CM}
\bsx_\cm\equiv \frac{1}{N}
\tr \bbx,\;\;\;\;  \bsp_\cm\equiv \tr \bbp,
\eea
%%%%%%%%%%%%%%%%%%%%%
\begin{figure}[t]
\begin{center}
\leavevmode
\epsfxsize=100mm
\epsfysize=50mm
\epsfbox{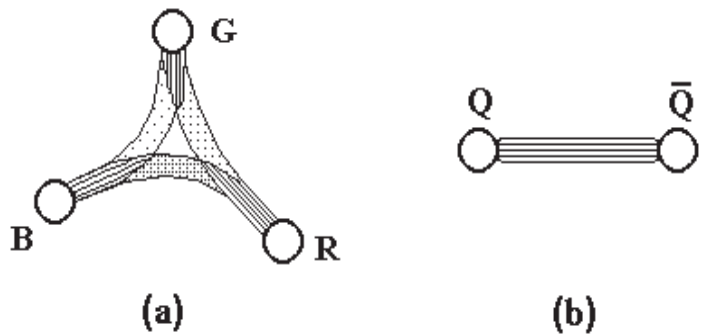}
\caption{{\it The net electric flux extracted from each
quark is equivalent in a baryon (a) and a meson (b).
The D0-brane--quark correspondence suggests the string-like
shape for flux inside a baryon (a).}}
\end{center}
\end{figure}
%%%%%%%%%%%%%%%%%%%%%
where we are using the convention $\tr {\bf 1}_N=N$. To specify the net
charge of a bound state (which is an extended object) its dynamics
should be
studied in zero magnetic and uniform electric fields, \ie$B_{ji}=0$ and
$E_i(\bbx,t)=E_{0i}$. \footnote{In a non-Abelian gauge theory a uniform
electric field can be defined up to a gauge transformation, which
sufficient for identification of white (singlet) states.} Since the fields are
uniform, they do not involve the $X^i$ matrices, and contain just the
U(1) part. In other words, under gauge transformations $E_{0i}$ and
$B_{ji}=0$ transform to $E'_i(\bbx,t)=U(\bbx,t)E_{0i}U^\dagger(\bbx,t)
=E_{0i}$ and $B'_{ji}= 0$. Thus the action (\ref{action5}) yields the
following equation of motion:
\bea
(Nm)\ddot{\bsx}_\cm=Nq {\bf E}_{0(1)},
\eea
in which the subscript (1) emphasizes the U(1) electric field. So the
c.m. interacts directly only with the U(1) of U($N$). From the
String Theory point of view, this observation is based on the simple fact
that the SU($N$) structure of D0-branes arises just from the internal
degrees of freedom inside the bound state. In other words,
the matrix behavior of the coordinates, and the resulted non-commutativity, is
just restricted to the relative positions of D0-branes. By this picture, we
may call this situation as `confined non-commutativity'
\cite{0108198,0104210,fat241,fat021}.
This behavior of D0-brane bound states is the same as that
of baryons. It means that each D0-brane feels the net effect of
other D0-branes as the white-complement of its color. In other
words, the field flux extracted from one D0-brane to the other ones
are the same as the flux between a color and an anti-color, Fig.1.
This shape for the electric flux are in agreement with the result
of field theory correlator method \cite{simonov}.
It was pointed that the gauge symmetry associated to gauge field
$(A_0(\bbx,t),\bba(\bbx,t))$, though looking similar to  the
non-Abelian gauge theories, is in intrinsic U(1). Based on the
observation we have made here about whiteness of the bound state, we may
argue in the phase that all of the observable states should have equivalent
amount of U($N$) sectors, the symmetry appears to be restricted, and
equivalently as U(1). In fact it is the case that we expect to see when
we are faced with matrix coordinates as relevant degrees of freedom.

It is desirable to assign a net charge different
from $Nq$ to the c.m. It can be done simply by modifying the action (\ref{action5})
\bea\label{action6}
S'[a_t,\bbx]=S[a_t,\bbx]+\int dt
\bigg(Nq'\dot{\bsx}_\cm\cdot \bba(\bsx_\cm,t)-Nq'A_0(\bsx_\cm,t)\bigg),
\eea
in which $S[a_t,\bbx]$ is the action (\ref{action5}). By this action
the charge of c.m. is equal to $N(q+q')$, rather than $Nq$.

Now, let us ignore for the moment the external
gauge field $(A_0,\bba)$. The equations of motion can be solved by
diagonal configurations, such as:
\bea
\bbx(t)&=&{\rm diag.} (\bsx_1(t),\cdots, \bsx_N(t)),\nonumber\\
a_t(t)&=&{\rm diag.} (a_{t1}(t),\cdots,a_{tN}(t)),
\eea
with $\bsx_a=\bsx_{a0}+\bsv_a t$, $a=1,\cdots,N$. By this configuration,
we restrict the U($N$) generators to the $N$ dimensional Cartan sub-algebra. This
configuration describes the ``classical" free motion
of $N$ D0-branes, neglecting the effects of the strings (and the symmetry
supported by them). Of course the situation is different when we consider
quantum effects, and consequently it will be realized that the dynamics of
the off-diagonal elements affect the dynamics of D0-branes significantly.
Concerning the effect of the strings, one may try to extract the effective
theory for D0-branes, \ie for the diagonal configurations. In particular,
it will be found out that the commutator potential is responsible for
the formation of the bound state, and by a simple dimensional analysis we
understand that the size of the bound state, $\ell$, is $\sim m^{-1/3} l^{2/3}$.
As in \cite{02414} (see also \cite{fat021,fat241}), let us take the
example of static D0-branes. For this configuration one can easily
calculate one-loop effective potential between the quarks, getting
\cite{fat021,fat241,02414}:
\bea\label{one-loop}
V_{{\rm one-loop}}\sim 4\pi\frac{d-1}{2}\sum_{a>b=1}^N
\frac{|\bsx_a-\bsx_b|}{l^2}.
\eea
This result shows the linear potential between each pair of D0-branes.
Previously we mentioned that, by qualitative considerations, what should
be the shape of the electric flux (Fig.1). Now, by interpretation of
(\ref{one-loop}) as the effective potential of a constituent quark model,
we are enable to know something more about the bound state and more quantitative
details. One can trace supports for the linear behavior of the
potential in the literature, namely results by lattice calculations \cite{bali}\cite{cornwall},
and things we expect from the spin-mass Regge trajectories.
In \cite{kirishna} by taking the linear potential between quarks of a
baryonic state in transverse direction of the light cone frame, the
structure functions are obtained in  good agreement with the observed
ones. Since the original theory is invariant under the rotation among the color indices
$1,\cdots, N$, we mention that only the states which are singlets under the
(global) rotation among the indices can be accepted as the physical states
of effective theory for diagonal elements.

The formulation we presented above is in the non-relativistic limit.
Though it is expected this limit produces good results for heavy quarks,
for light or massless quarks we should change our approach. One way
can be starting by a covariant theory; treating time and space
equivalently. In this way, although the terms responsible for kinetic
energy and interaction with external gauge fields find reasonable forms
(see \cite{0104210,0108198}), the main problem will appear to be
with potentials as $[X^\mu,X^\nu]^2$. Instead one may follow another
approach to say something about the covariant theory. The world-line
formulation we have here is that of the M(atrix) model conjecture,
accompanied with the interaction terms with external gauge fields.
For the case of the dynamics of a massless charged particle with ordinary
coordinates, we can see easily that the light-cone dynamics have a form
similar to that we have in action (\ref{action5}); see Appendix of
\cite{fat021}. To approach the covariant formulation, following
finite-$N$ interpretation of \cite{9704080}, it is reasonable to interpret
things in the DLCQ framework \cite{05484,fat021,fat241,0108198}.
In this way of interpretation, the mass parameter $m$ is the longitudinal
momentum, and the spatial directions present the transverse coordinates in
the light-cone frame. In addition, according to the specific form of action
(\ref{action5}) the rest mass of quraks is assumed to be zero (see \cite{fat021,fat241}).

In \cite{05484,fat021} and \cite{0108198} the problem of scattering of
1) a D0-brane off another one and, 2) a D0-brane bound state off an external gauge
field probe, were considered. As we mentioned  above,
both of the scattering processes can be interpreted in the light-cone
frame. For the case of scattering of a D0-brane off
another one, the expectations for the well known Regge behavior
are satisfied. As for the problem of interaction between D0-brane
bound state and `photons' of gauge field, the interesting
observations is expected for the regime in which the details
of the bound state can be probed. Here we just
present the general expected features; see \cite{0108198} for more details.
As argued in before, the
external field depend on the internal coordinates of the bound state
under the symmetrization condition in matrix coordinates.
One way to cover the symmetrization is to use the so-called
`non-Abelian Fourier expansion' \cite{0108198}.
For an arbitrary function $f(\bbx,t)$ the non-Abelian Fourier expansion will be found to be:
\bea\label{NFE}
f(\bbx,t)= \int d\bsk\; \bar f (\bsk,t)\;\e^{i\bsk\cdot\bbx},
\eea
in which $\bar f(\bsk,t)$ are the Fourier components of the function
$f(\bsx,t)$ (\ie function by ordinary coordinates) which is defined by the
known expression:
\bea
\bar f (\bsk,t) \equiv \frac{1}{(2\pi)^d}\int d\bsx\; f(\bsx,t)\; \e^{-i\bsk\cdot\bsx}.
\eea
Since the momentum numbers $k_i$'s are ordinary numbers, and so commute
with each other, the symmetrization prescription is automatically
recovered in the expansion of the momentum eigen-functions $\e^{i\bsk\cdot
\bbx}$. Now, by using the symmetric expansion (\ref{NFE}), we can imagine some
general aspects of the interaction between D0-brane bound states and RR
photons. As we mentioned in before the size of the bound state, for finite number $N$ of D0-branes is
finite and is of order of $\ell\sim m^{-1/3} l^{2/3}$.

Before proceeding further, we should distinguish the dynamics of the c.m.
from the internal degrees of freedom of the bound state. As mentioned in
before, the c.m. position and momentum of the bound state are presented by
the $U(1)$ sector of the $U(N)=SU(N)\times U(1)$, and thus the information
related to the c.m. can be gained simply by the $\tr$operation.
So, the internal degrees of freedom of the bound state, which
consist the relative positions of $N$ D0-branes together with the dynamics
of strings stretched between D0-branes, are described by the $SU(N)$
sector of the matrix coordinates. It is easy to see that the commutator
potential in the action has some flat directions, along which the
eigen-values can take arbitrary large values. But it is understood that,
by considering the quantum effects and in the case that we expect
formation of the bound state, we should expect suppression the large
values of the internal degrees of freedom \cite{dewit-nicolai}.
Consequently, it is expected that the $SU(N)$ sector of matrix coordinates
take mean values like $\langle X^i_\alpha \rangle \sim \ell$ ($\alpha=1,\cdots,
N^2-1$, not $\alpha=0$ as c.m.), with $\ell$ as the bound state size scale
mentioned in above. We should mention that, though
the c.m. is represented by the $U(1)$ sector, but its dynamics is affected
by the interaction of the ingredients of bound state with the $SU(N)$
sector of external fields, similar to the situation we imagine in the case
of the Van der Waals force.

The important question about the interaction of a bound state (as an
extended object) with an external field, is about `the regime in which the
substructure of bound state is probed'. As we mentioned in introduction,
in our case the quanta of RR fields are the representatives of the
external field. The quanta are coming from a `source' and so, as it makes
easier things, we ignore its dynamics. The source is introduced to our
problem by the gauge field $A_\mu(\bsx,t)$. These fields appear in the action
by functional dependence on matrix coordinates $\bbx$'s.  In fact this is the
key of how we can probe the substructure of the bound state. According to
the non-Abelian Fourier expansion we mentioned in above, we have
\bea
A_\mu(\bbx,t)= \int d\bsk\; \bar A_\mu (\bsk,t)\e^{i\bsk\cdot\bbx},
\eea
in which $\bar A_\mu(\bsk,t)$ is the Fourier components of the fields
$A_\mu(\bsx,t)$ (\ie fields by ordinary coordinates). One can imagine the
scattering processes which are designed to probe inside the bound state.
Such as every other scattering process two limits of momentum modes,
corresponding to long and short wave-lengths, behave differently.

%%%%%%%%%%%%%%%%%%%%%%%%%%%%%%%%%%%%%%%%%%%%%%%%%%%%%%%%%%%%%%%%
\begin{figure}[t]
\begin{center}
\leavevmode
\epsfxsize=80mm
\epsfysize=50mm
\epsfbox{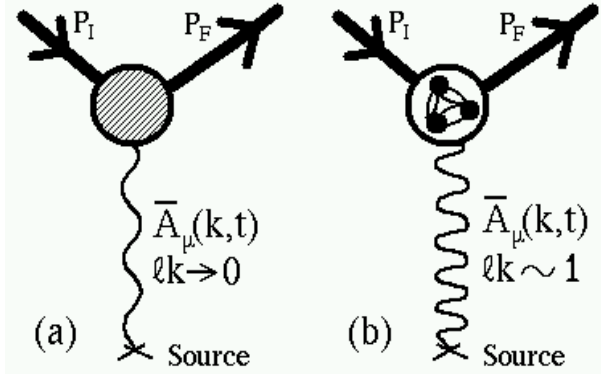}
\caption{{\it Substructure is not seen by the long wave-length modes (a). Due
to functional dependence on matrix coordinates, the short wave-length
modes can probe inside the bound state (b). $\ell$ and $\bar A_\mu (k,t)$
represent the size of the bound state and the Fourier modes,
respectively.}}
\end{center}
\end{figure}
%%%%%%%%%%%%%%%%%%%%%%%%%%%%%%%%%%%%%%%%%%%%%%%%%%%%%%%%%%%%%%%%

In the limit $\ell |\bsk|\goes 0$ (long wave-length regime), the field
$A_\mu$ is not involved by $\bbx$ matrices mainly. It means that the fields
appear to be nearly constant inside the bound state, and in rough
estimation we have
\bea
\e^{i\bsk\cdot\bbx} \sim \e^{i\bsk\cdot\bbx_\cm}.
\eea
So in this limit we expect that the substructure and consequently
non-commutativity will not be seen; Fig.-2a. As the consequence, after
interaction with a long wave-length mode, it is not expected that the
bound state jump to another energy level different from the first one. It
should be noted that the c.m. dynamics can be affected as well in this
case.

In the limit $\ell |\bsk|=$finite (short wave-length regime), the fields
depend on coordinates $\bbx$ inside the bound state, and so the substructure
responsible for non-commutativity should be probed; Fig.-2b. In fact, we
know that the non-commutativity of D0-brane coordinates is the consequence
of the strings which are stretched between D0-branes. In this case,
it is completely expectable that the energy level of the incoming and
outgoing bound states will be different, since the ingredients of bound
state substructure can absorb quanta of energy from the incident wave. In
this case the c.m. dynamics can be affected in a novel way by the
interaction of the substructure with the external fields (the Van der
Waals effect). In general case, one can gain more information about the substructure of a
bound state by analysing the `recoil' effect on the source. To do this,
one should be able to include the dynamics of the source in the
formulation. Considering the dynamics of source, in the terms of quantized
field theory, means that we consider the processes in which the source and
the target exchange `one quanta of gauge field' with definite wave-length
and frequency, though off-shell, as $A_\mu(\bsx,t) \sim \epsilon_\mu
\e^{i\bsk\cdot\bsx-i\omega t}$.

Up to now, we have considered things for the theory with one kind of
flavor. It is interesting to think about the case with more than one
flavor. One suggestion can be as follows: assume the flavor $A$ with mass $m_A$
is represented by the state $|\Psi_A(t)\rangle$. We may re-scale the
states as $|\Psi_A\rangle\goes |\tilde{\Psi}_A\rangle =(m_A)^{1/4}
|\Psi_A\rangle$. For a baryon consisting $N$ heavy flavors we define the
matrix coordinate as
\bea
\tilde{\bbx}(t)\equiv \left(\!\!\!
\begin{array}{cccc}
\langle\tilde{\psi}_1(t)|\hat{\bsx}|\tilde{\psi}_1(t)\rangle &
\langle\tilde{\psi}_2(t)|\hat{\bsx}|\tilde{\psi}_1(t)\rangle &
\dots & \langle\tilde{\psi}_N(t)|\hat{\bsx}|\tilde{\psi}_1(t)\rangle\\
\langle\tilde{\psi}_1(t)|\hat{\bsx}|\tilde{\psi}_2(t)\rangle & \dots & \dots& \dots\\
\vdots & \ddots & \ddots & \vdots\\
\langle\tilde{\psi}_1(t)|\hat{\bsx}|\tilde{\psi}_N(t)\rangle & \dots & \dots &
\langle\tilde{\psi}_N(t)|\hat{\bsx}|\tilde{\psi}_N(t)\rangle
\end{array}\!\!\!\right).
\eea
For this coordinate we take the action
\bea
S[\tilde{\bbx}]=\int dt \tr \bigg( \haf \dot{\tilde{\bbx}}\cdot \dot{\tilde{\bbx}} -
\cdots\bigg).
\eea
Now, for the well separated states, for which we have diagonal coordinates, the action
in terms of original coordinates (\ie before re-scaling) becomes
\bea
S[\bsx_A]=\int dt \sum_{A=1}^{N}\bigg( \haf m_A \dot{\bsx}_A\cdot \dot{\bsx}_A
- \cdots \bigg),
\eea
in which we see that each flavor has the expected kinetic term.
It is worth recalling that due to the color symmetry we expect,
the coordinate to which the symmetry transformation should apply is
$\tilde{\bbx}$.

In \cite{0104210} a conceptual relation between use of matrix coordinate
for non-Abelian gauge theory purposes and the
ideas concerned in special relativity is mentioned; see also
\cite{fat241,fat021,02414}. According to an interpretation of the special
relativity, it is meaningful if the `coordinates' and the
`fields' in a theory have some kinds of similar characters.
As an example, we observe that both the space-time coordinates $x^\mu$ and
the electro-magnetic potentials $A^\mu(x)$ transform equivalently (\ie as a
$(d+1)$-vector) under the boost transformations. Also by this way of
interpretation, the super-space formulations of supersymmetric field and
superstring theories are the natural continuation of the special
relativity program. In the case of use of matrix coordinates, it may be argued
that the relation between `matrix coordinates' and `matrix fields' (gauge
fields of a non-Abelian gauge theory) is one of the expectations which is
supported by the spirit of the special relativity. From the previous
discussion we recall, 1) the matrix character of gauge fields is the result
of dependence of them on matrix coordinates \cite{0104210}, 2) the symmetry
transformations of gauge fields is induced by the transformations of matrix
coordinates \cite{0104210}, 3) the transformations of fields in the
theory on matrix space appeared to be similar to those of non-Abelian
gauge theories, relations (\ref{NAT}) and (\ref{gtfs}). This way of
interpretation leads us to conclude that the non-Abelian gauge fields in a confined
theory do not have an independent character, and they are introduced to
the formalism due to functional dependence on the matrix coordinates of `bounded
quarks'. It seems very interesting when we note that by the present situation of
experimental data, the existence of pure gluonic states, the so-called glueballs,
is quite doubtful. This lack of detection may be taken as a support for the
interpretation presented above.

%%%%%%%%%%%%%%%%%%%%%%%%%%%%%%%%%%%%%%%%%%%%%%%%%%%%%%%%%%%%%%%%
\vspace{.5cm}

{\bf Acknowledgement:} The author is grateful to A. Shariati, and specially to
M. Khorrami for helpful discussions. The comments on the
manuscript by Gh. Exirifard, and specially by
S. Parvizi and M.M. Sheikh-Jabbari are acknowledged.

%%%%%%%%%%%%%%%%%%%%%%%%%%%%%%%%%%%%%%%%%%%%%%%%%%%%%%%%%%%%%%%%


\begin{thebibliography}{99}
\bibitem{werner} W. Heisenberg, ``Quantum-Theoretical Re-Interpretation
Of Kinematic And Mechanical Relations," (in German) Zs. Phys. {\bf 33} (1925) 879.

\bibitem{02414} A.H. Fatollahi, ``Do Quarks Obey D-Brane Dynamics?,"
Europhys. Lett. {\bf 53(3)} (2001) 317, hep-ph/9902414.

\bibitem{05484} A.H. Fatollahi, ``Do Quarks Obey D-Brane Dynamics? II,"
Europhys. Lett. {\bf 56(3)} (2001) 523, hep-ph/9905484.

\bibitem{fat021} A.H. Fatollahi, ``D0-Branes As Light-Front Confined
Quarks," Eur. Phys. J. {\bf C19} (2001) 749, hep-th/0002021.

\bibitem{fat241} A.H. Fatollahi, ``D0-Branes As Confined Quarks," talk given
at ``Isfahan String Workshop 2000, May 13-14, Iran,"
hep-th/0005241.

\bibitem{9510017} J. Polchinski, ``Dirichlet-Branes And Ramond-Ramond
Charges," Phys. Rev. Lett. {\bf 75} (1995) 4724, hep-th/9510017.

\bibitem{tasi} J. Polchinski, ``TASI Lectures On D-Branes," hep-th/9611050.

\bibitem{9510135} E. Witten, ``Bound States Of Strings And p-Branes,"
Nucl. Phys. {\bf B460} (1996) 335, hep-th/9510135.

\bibitem{pol-book} J. Polchinski, ``String Theory," vol. I, Cambridge Univ. Press,
pp. 184 and 268.

\bibitem{0103262} A.H. Fatollahi, ``On Non-Abelian Structure From Matrix
Coordinates," Phys. Lett. {\bf B512} (2001) 161, hep-th/0103262.

\bibitem{0104210} A.H. Fatollahi, ``Electrodynamics On Matrix Space:
Non-Abelian By Coordinates," Eur. Phys. J. {\bf C21} (2001) 717,
hep-th/0104210.

\bibitem{0108198} A.H. Fatollahi, ``Interaction Of D0-Brane Bound
States And Ramond-Ramond Photons," Phys. Rev. {\bf D65} (2002) 046004,
hep-th/0108198.

\bibitem{0007023} A.H. Fatollahi, ``Gauge Symmetry As Symmetry Of Matrix
Coordinates," Eur. Phys. J. {\bf C17} (2000) 535, hep-th/0007023.

\bibitem{9910053} R.C. Myers, ``Dielectric-Branes," JHEP {\bf 9912} (1999)
022, hep-th/9910053; W. Taylor and M. Van Raamsdonk, ``Multiple Dp-Branes In
Weak Background Fields," Nucl. Phys. {\bf B573} (2000) 703,
hep-th/9910052.

\bibitem{simonov} D.S. Kuzmenko and Y.A. Simonov,
``QCD String In Mesons And Baryons,"
Phys. Atom. Nucl. {\bf 64} (2001) 107; Yad. Fiz. {\bf 64} (2001) 110,
hep-ph/0010114; A.D. Giacomo, H.G. Dosch, V.I. Shevchenko and Y.A. Simonov,
``Field Correlators In QCD. Theory And Applications,"
hep-ph/0007223.

\bibitem{bali} G.S. Bali, ``QCD Forces And Heavy Quark Bound States,"
Phys. Rept. {\bf 343} (2001) 1, hep-ph/0001312, page 73;
C. Alexandrou, Ph. de Forcrand and A. Tsapalis, ``The Static Baryon
Potential," nucl-th/0111046.

\bibitem{cornwall} J.M. Cornwall, ``Baryon Wilson Loop Area Law In QCD,"
Phys. Rev. {\bf D54} (1996) 6527.

\bibitem{kirishna} G.S. Kirishnaswami, ``A Model Of Interacting
Partons For Hadronic Structure Functions," hep-ph/9911538; G.S.
Kirishnaswami and S. G. Rajeev,
Phys. Lett. {\bf B441} (1998) 449.

\bibitem{9610043} T. Banks, W. Fischler, S.H. Shenker and L. Susskind, ``M
Theory As A Matrix Model: A Conjecture," Phys. Rev. {\bf D55}
(1997) 5112, hep-th/9610043.

\bibitem{review-matrix} T. Banks, ``Matrix Theory," Nucl. Phys. Proc.
Suppl. {\bf 67} (1998)  180, hep-th/9710231; ``TASI Lectures On
Matrix Theory," hep-th/9911068; D. Bigatti and L. Susskind,
``Review Of Matrix Theory," hep-th/9712072.

\bibitem{9704080} L. Susskind, ``Another Conjecture About M(atrix)
Theory,"  hep-th/9704080.

\bibitem{dewit-nicolai} B. de Wit, ``Supersymmetric Quantum Mechanics,
Supermembranes And Dirichlet Particles," Nucl. Phys. Proc. Suppl. {\bf
B56} (1997) 76, hep-th/9701169; H. Nicolai and R. Helling,
``Supermembranes And M(atrix) Theory," hep-th/9809103; B. de Wit,
``Supermembranes And Super Matrix Models,"  hep-th/9902051.

\end{thebibliography}
\end{document}